\documentclass[11pt]{article}
\usepackage{lipsum} 
\usepackage{datetime}

\usepackage[utf8]{inputenc}
\usepackage{hyperref}
\usepackage{setspace}
\usepackage{palatino}
\usepackage{graphicx}
\usepackage{float}
\usepackage{titling} 
\usepackage{multirow}
\usepackage{lscape}
\usepackage{amsmath}
\usepackage{amssymb}
\usepackage{xspace}
\usepackage{subcaption}
\usepackage[a4paper, total={6in, 9.5in}]{geometry}
\fontfamily{ppl}\selectfont 

\usepackage[american]{babel}

\usepackage{setspace}

\newcommand{\pt}{\ensuremath{p_{\mathrm{T}}}\xspace}
\newcommand{\ptmiss}{\ensuremath{\pt^\text{miss}}\xspace}
\newcommand{\ma}{\ensuremath{m_{\textrm{a}}}\xspace}
\newcommand{\mA}{\ensuremath{m_{\textrm{A}}}\xspace}
\newcommand{\sintheta}{\ensuremath{\sin\theta}\xspace}
\newcommand{\tanbeta}{\ensuremath{\tan\beta}\xspace}

\title{Shedding Light on Dark Matter via the Higgs portal}
\author{Presented at the 32nd International Symposium on \\ Lepton Photon Interactions at High Energies,\\
Madison, Wisconsin, USA\\\\\\
Shivani Lomte \\ University of Wisconsin-Madison \\ {\it On behalf of the CMS Collaboration}
}    
\date{August 25-29, 2025}

\begin{document}

\maketitle
\selectlanguage{american}
\begin{abstract}
A selection of new results from the CMS experiment that probe Higgs-portal frameworks are presented. These searches target scenarios in which dark matter particles are produced in association with the Standard Model Higgs boson. No significant deviations from the Standard Model predictions are observed, and new limits are set using the Run-2 proton-proton collision dataset at $\sqrt{s}=13$ TeV. 
\end{abstract}

\selectlanguage{american} 

\section{Introduction}
The standard model (SM) of particle physics describes the measurements performed by high-energy physics experiments with an unprecedented precision. However, it does not incorporate several important phenomena, such as the nature of dark matter and the matter-antimatter asymmetry in the Universe. The Beyond the SM (BSM) physics program at the ATLAS and CMS experiments at CERN LHC are searching for new particles to shed light on these unanswered questions. 

If dark matter (DM) particles are produced in high energy proton-proton collisions at the LHC, their presence can be inferred from the missing transverse momentum ($\ptmiss$). Mono-$X$ searches target events where a visible SM particle ($X$) recoils against the invisible DM, producing a distinctive $X + \ptmiss$ signature. The SM particle $X$ can be emitted as an initial state radiation (ISR) or as part of a new coupling of DM to the SM. Higgs boson ($h$) production via ISR is suppressed due to the mass dependence of its coupling strength to fermions and its loop-suppressed coupling to gluons. Therefore, the $h$ + DM production can directly probe the effective DM--SM coupling. 

In this report, three latest results from the CMS experiment~\cite{cmspaper} are highlighted based on data recorded at $\sqrt{s}=13$ TeV during Run-2 of the LHC. Two of these searches involve the SM Higgs and one involves a dark Higgs produced in association with DM particles. 

\section{New Physics Results}
\subsection{Search for mono Higgs ($b\bar{b}$)}
In this analysis~\cite{pas-sus-24-007}, the signal events are characterized by large $\ptmiss$, no isolated lepton ($e$, $\mu$, or $\tau$) or photon, and the presence of a Higgs boson candidate. Two reconstruction strategies are employed: a resolved category for low to moderate $\pt$ Higgs boson candidate and a boosted category to capture high $\pt$. In the resolved category, each b-jet is reconstructed as an anti-$k_{T}$ jet with distance parameter $R=0.4$, and b-tagged with the $\textsc{DeepJet}$ algorithm. The $b\bar{b}$ dijet system is required to have $\pt>200$ GeV and $70<m_{b\bar{b}}<160$ GeV. In the boosted category, the decay products are highly collimated and reconstructed into a single anti-$k_{T}$ jet with distance parameter $R=0.8$, with $\pt>200$ GeV and $70<m_{\textrm{SoftDrop}}<160$ GeV. The analysis sensitivity to boosted signatures is enhanced by means of improved identification of $h \rightarrow b\bar{b}$ using a DNN based tagging algorithm, $\textsc{ParticleNet}$. 

\begin{figure}[!h]
\begin{minipage}{0.50\linewidth}
\centerline{\includegraphics[width=1.1\linewidth]{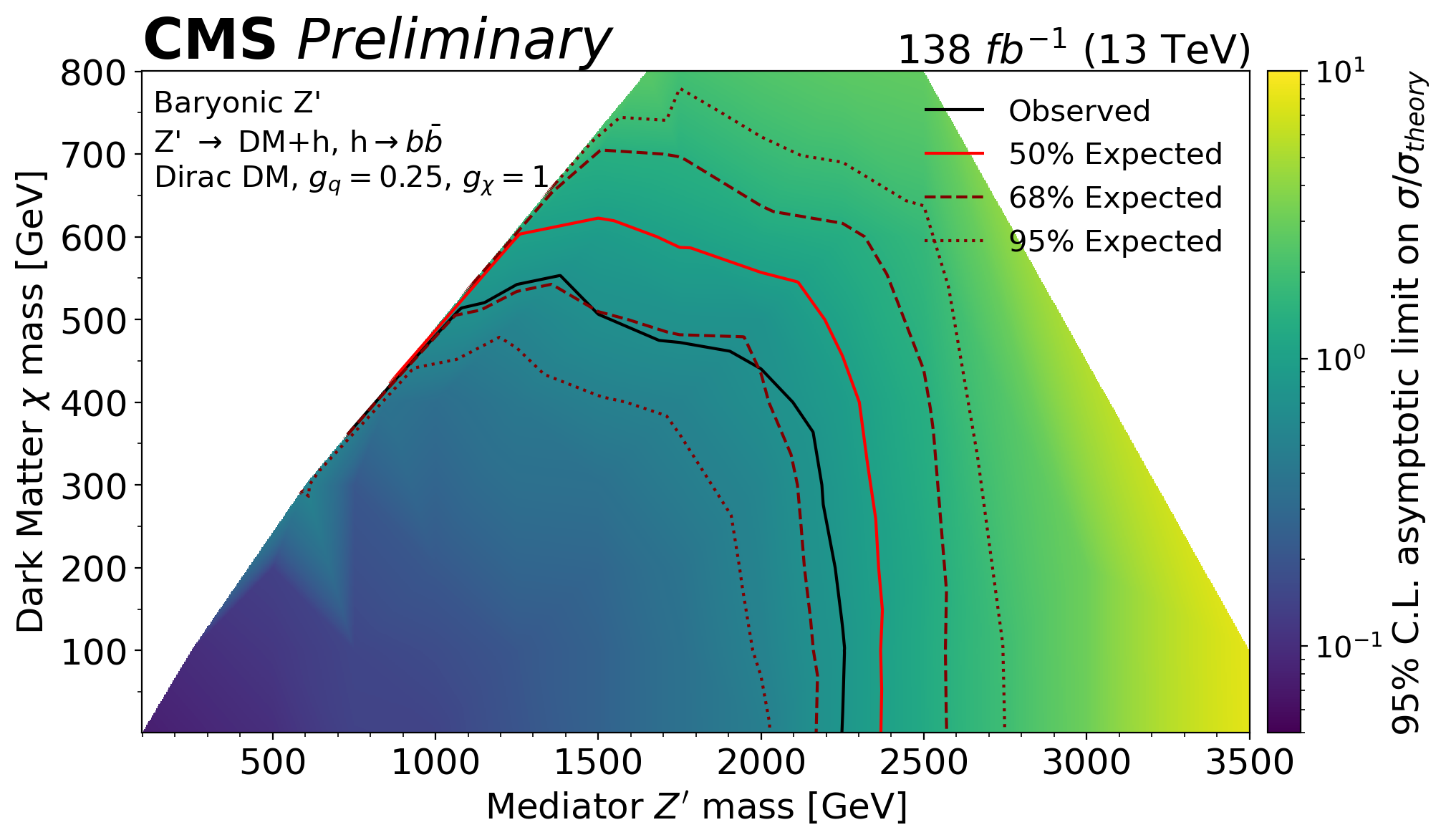}}
\end{minipage}
\begin{minipage}{0.55\linewidth}
\centering
\includegraphics[width=0.45\linewidth]{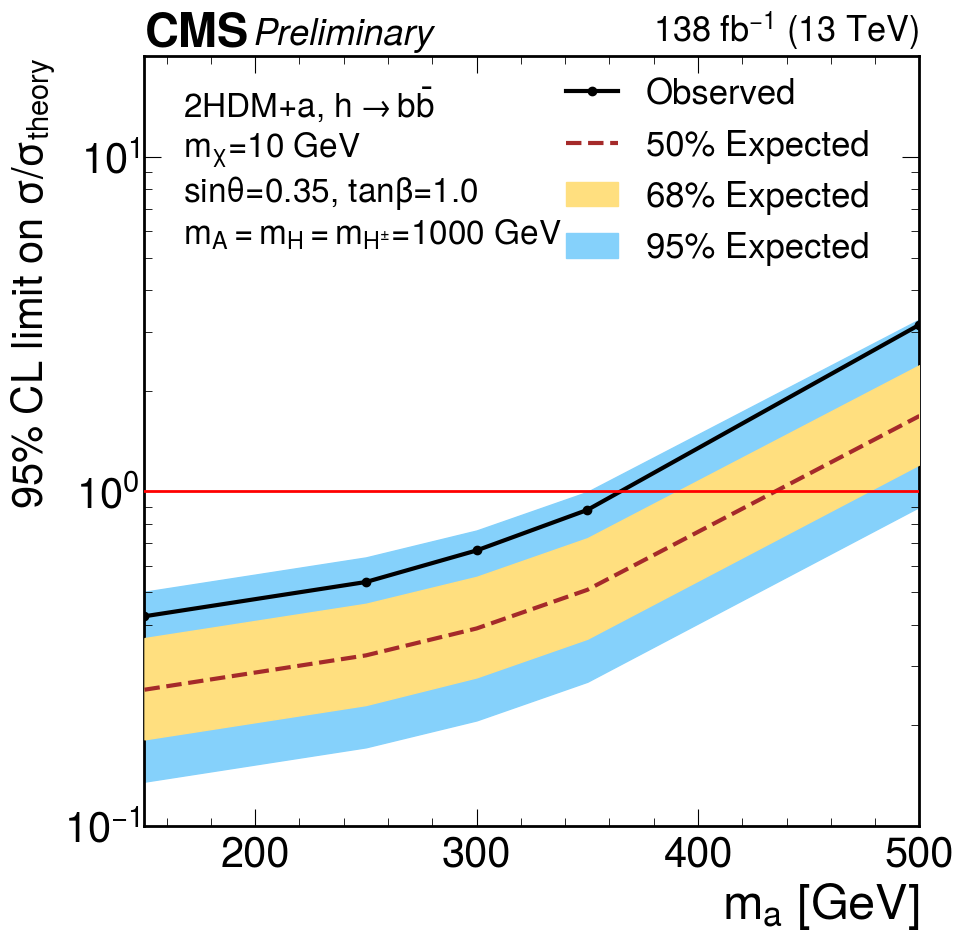}
\includegraphics[width=0.45\linewidth]{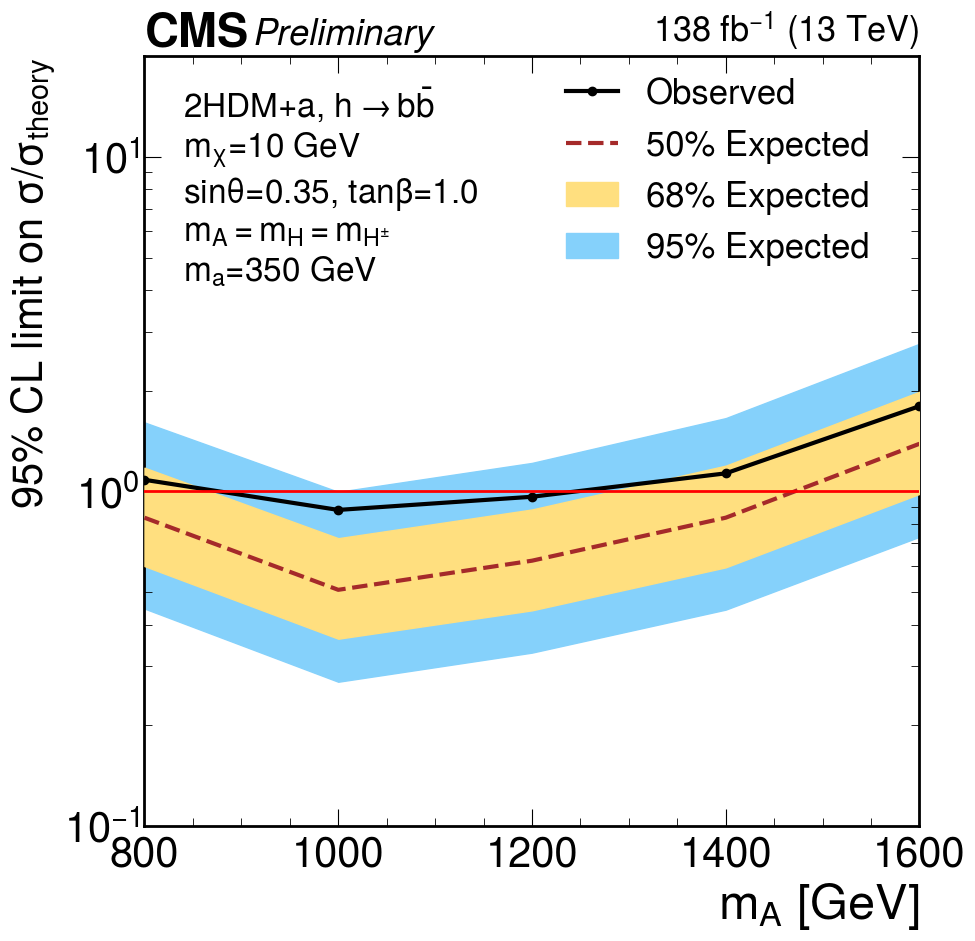}
\includegraphics[width=0.45\linewidth]{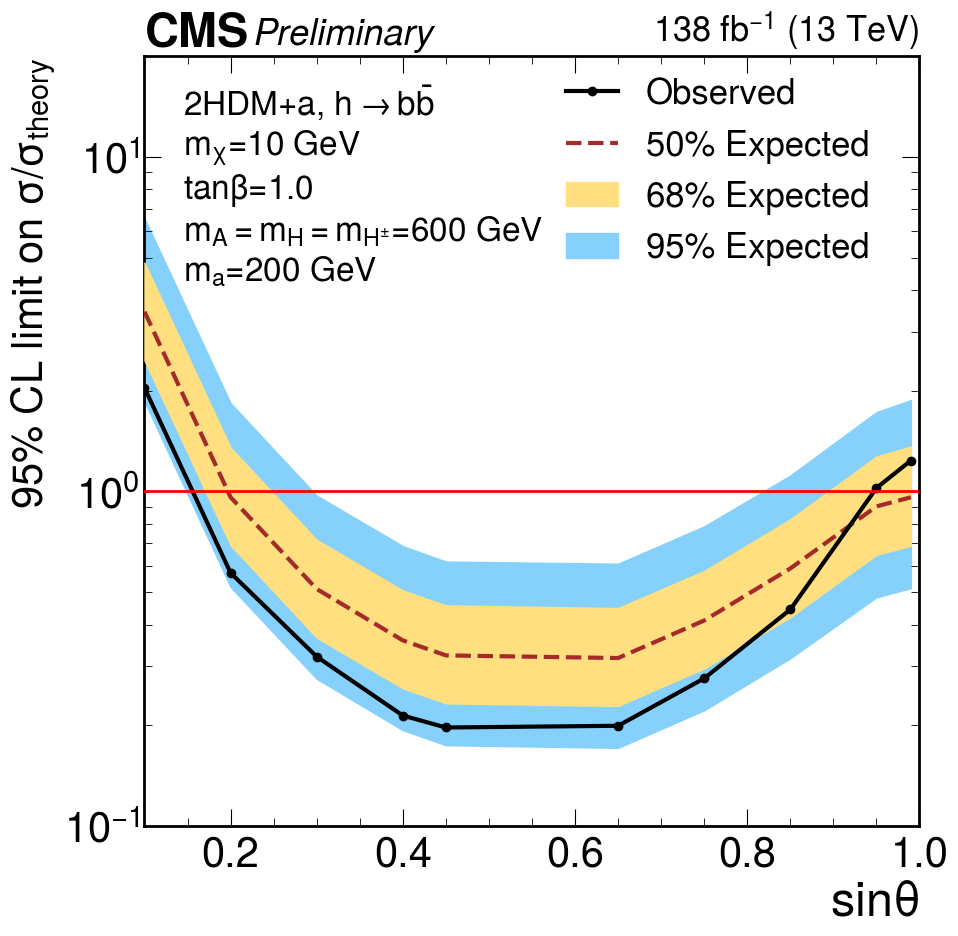}
\includegraphics[width=0.45\linewidth]{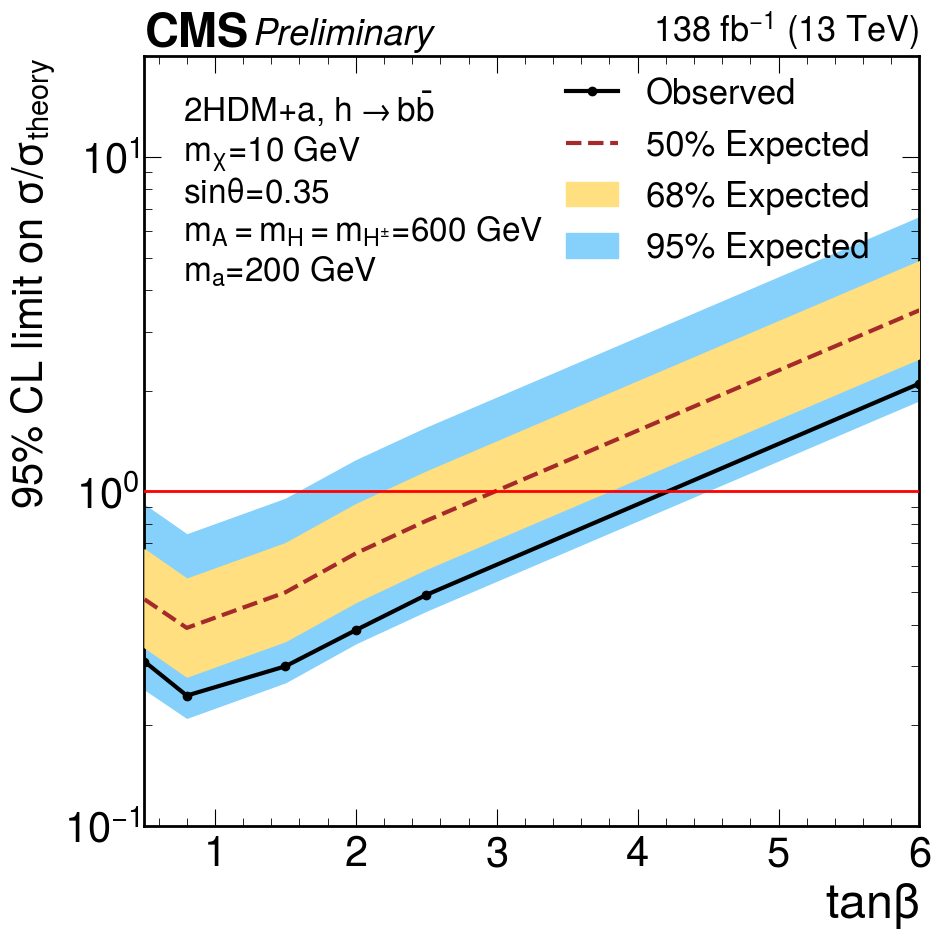}
\end{minipage}
\caption[]{The 95\% CL upper limit on signal strength $\mu$ for baryonic-$\textrm{Z}^{\prime}$ (left) in the ($m_{\textrm{Z}^{\prime}}$, $m_{\chi}$) plane and for 2HDM+a (right) as a function of model parameters: $\ma$, $\mA$, $\sintheta$, and $\tanbeta$ while fixing the values of other parameters, as indicated in the figures~\cite{pas-sus-24-007}.}
\label{fig:monoHiggs_bb}
\end{figure}

A combined maximum likelihood fit is performed with the observables $\ptmiss$ and Higgs candidate mass, $m_{\textrm{SoftDrop}}$ in the boosted category and $m_{b\bar{b}}$ in the resolved category. The data agree with the SM prediction, and no significant excess is observed. The results are interpreted into two simplified benchmark models: a baryonic-$\textrm{Z}^{\prime}$ model and a two-Higgs-doublet model with an additional pseudoscalar (2HDM+a). The baryonic-$\textrm{Z}^{\prime}$ model predicts a vector boson mediator $\textrm{Z}^{\prime}$ which radiates a Higgs boson and then decays into a pair of DM particles, $\chi \bar{\chi}$. The coupling strength of $\textrm{Z}^{\prime}$ particle to SM quarks is fixed to 0.25, and to DM particles is set to 1. The free parameters in this model are the $\textrm{Z}^{\prime}$ and DM masses, which are varied in this search. Figure~\ref{fig:monoHiggs_bb} (left) shows the upper limit in a two-dimensional scan of $m_{\textrm{Z}^{\prime}}$ and $m_{\chi}$ for the baryonic-$\textrm{Z}^{\prime}$ model with observed (black) and expected (red) exclusion regions. 

For the 2HDM+a model, the production cross section and kinematic distributions are governed by model parameters, the light and heavy pseudoscalar masses - $\ma$ and $\mA$, the sine of the mixing angle $\sintheta$, and the ratio of the vacuum expectation values $\tanbeta$. The DM particle mass is assumed to be 10 GeV. Figure~\ref{fig:monoHiggs_bb} (right) shows the exclusion limit on signal strength modifier at 95\% CL for each of the four one-dimensional scans - $\ma$, $\mA$, $\sintheta$, and $\tanbeta$. In each of these parameter scans, the remaining three parameter values are fixed as indicated in the figures.

\subsection{Search for mono Higgs ($\tau\tau$)}
This search~\cite{pas-sus-23-012} focuses on tau lepton pair ($\tau\tau$) final states with the highest branching fractions, $h\rightarrow \tau\tau \rightarrow e \tau_{h}, \ \mu \tau_{h}, \ \tau_{h}\tau_{h}$. Despite the lower branching ratio of $h \rightarrow \tau\tau$ compared to the dominant $h \rightarrow b\bar{b}$, this channel takes advantage of the smaller SM background contribution. Lepton triggers are used which have lower thresholds than the $\ptmiss$ triggers, allowing to probe DM scenarios with low $\ptmiss$ signatures. Thus, the $b\bar{b}$ and $\tau\tau$ modes serve as complementary channels. 

The hadronically decaying tau lepton ($\tau_{h}$) candidate is tagged using the $\textsc{DeepTau}$ algorithm. For all $\tau$ pair categories, $\ptmiss>105$ GeV is required, with the visible $\pt>65$ GeV and visible mass $<125$ GeV of the $\tau\tau$ system to ensure it is compatible with a SM Higgs boson. This analysis employs a strategy based on control samples in data, known as the “misidentification factor (MF) method”. This collectively estimates background events with quark- or gluon-jets misidentified as $\tau_{h}$, instead of measuring this background separately for W+jets and QCD events.

\begin{figure}[!h]
\begin{minipage}{0.50\linewidth}
\centerline{\includegraphics[width=1.1\linewidth]{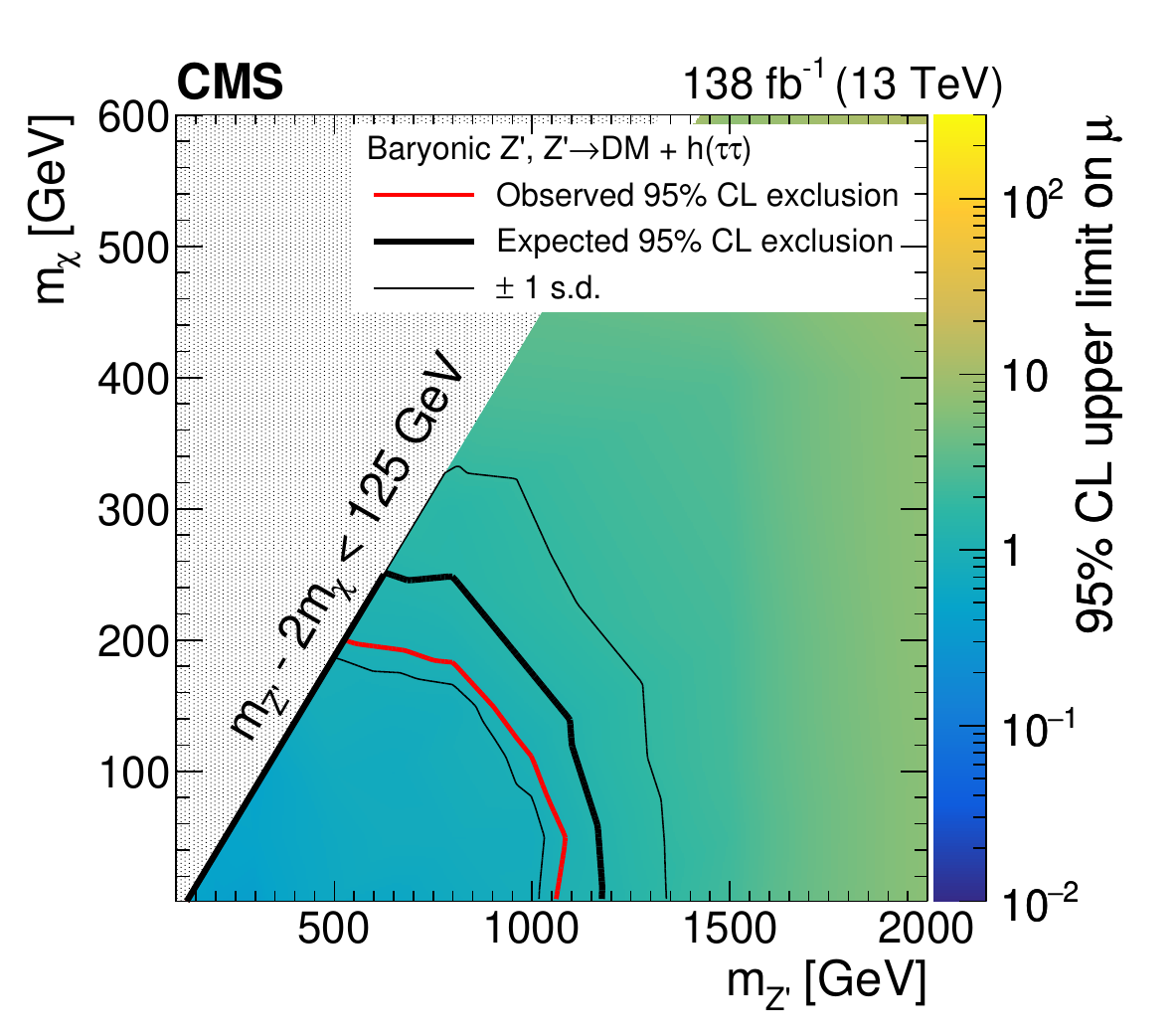}}
\end{minipage}
\begin{minipage}{0.55\linewidth}
\centering
\includegraphics[width=0.45\linewidth]{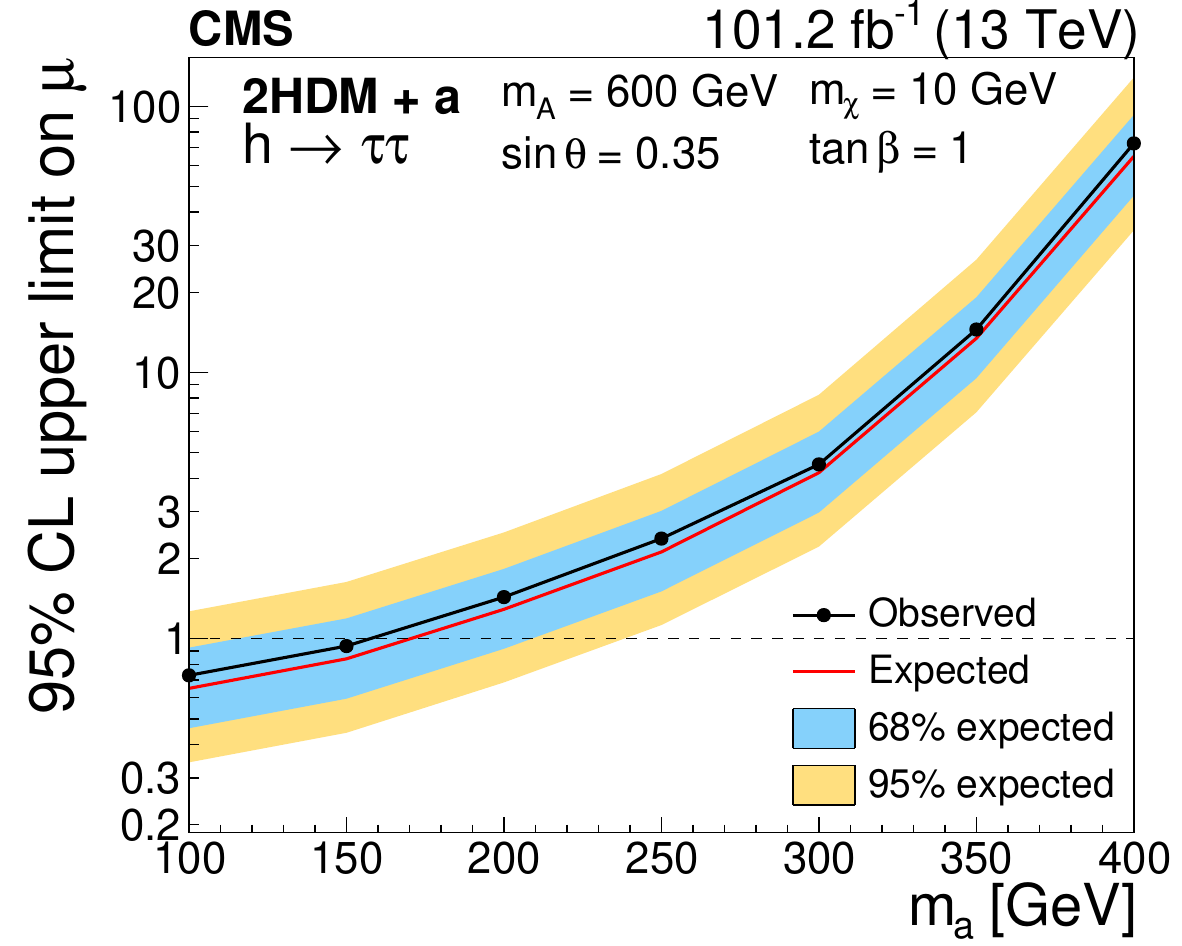}
\includegraphics[width=0.45\linewidth]{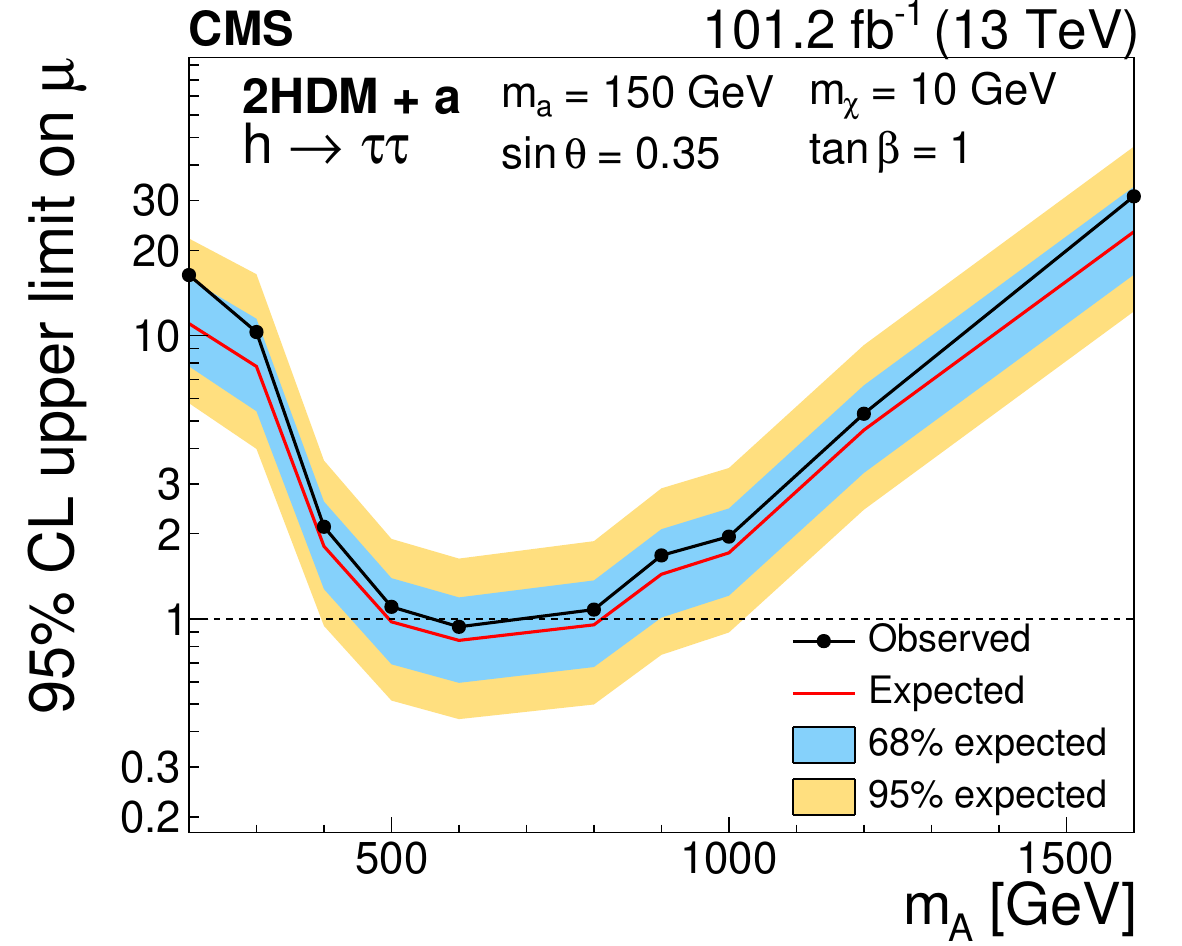}
\includegraphics[width=0.45\linewidth]{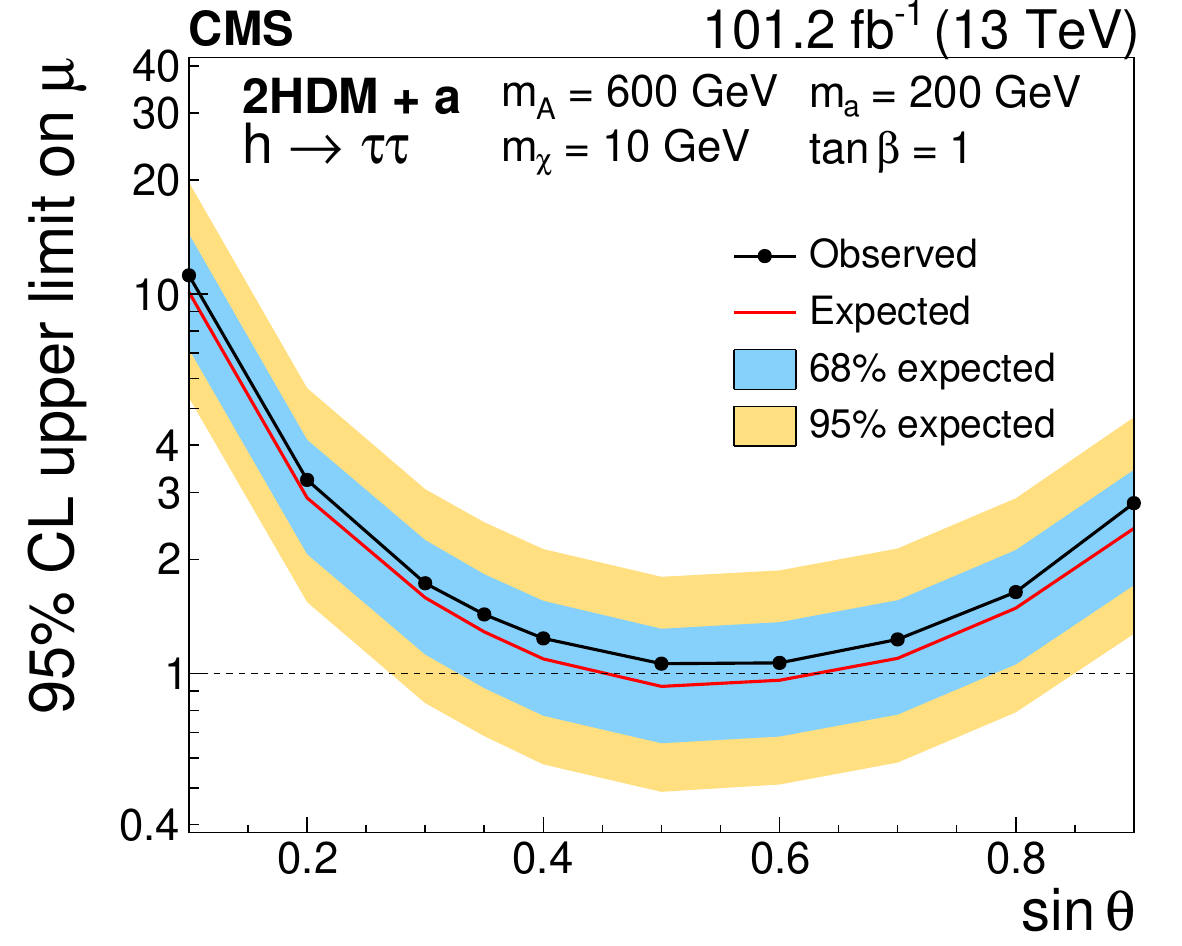}
\includegraphics[width=0.45\linewidth]{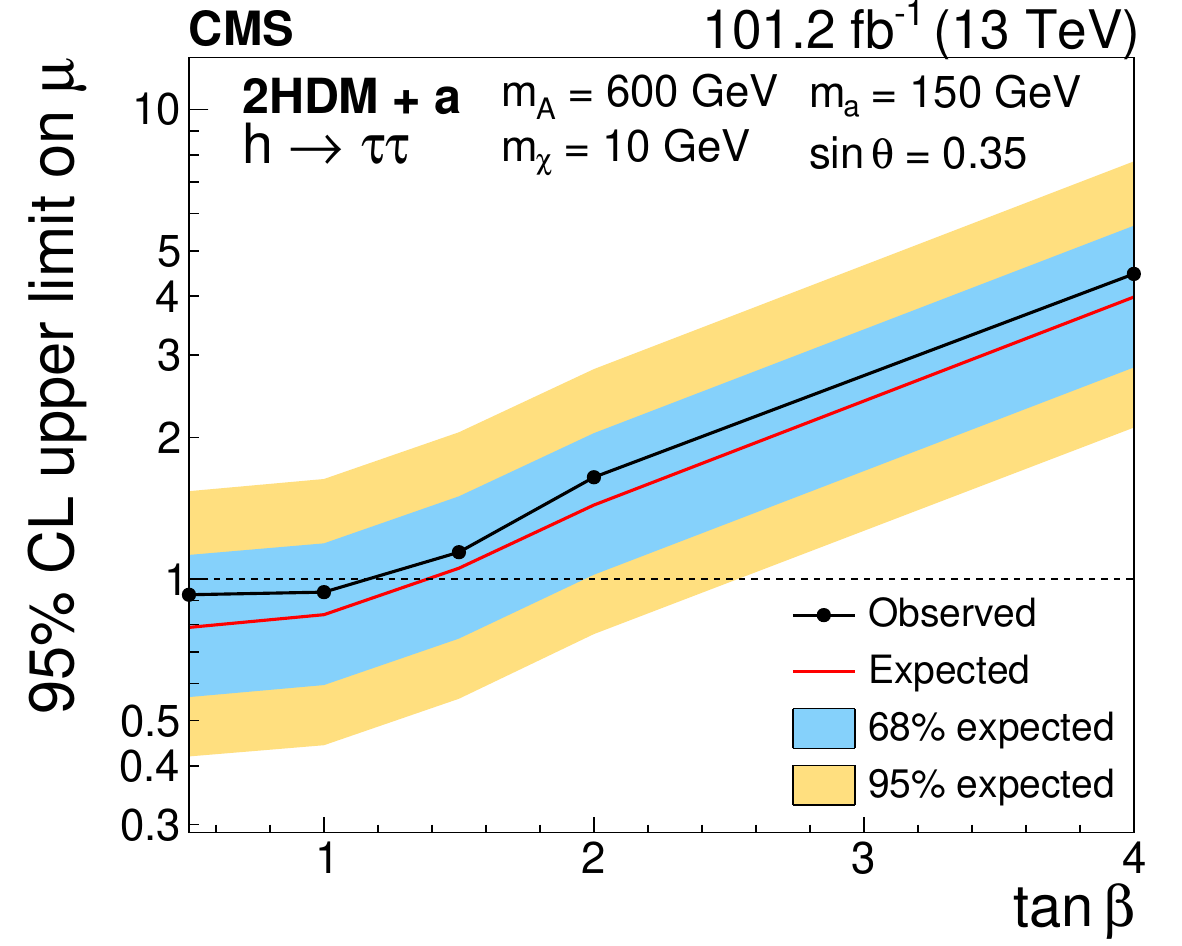}
\end{minipage}
\caption[]{The 95\% CL upper limit on signal strength $\mu$ for baryonic-$\textrm{Z}^{\prime}$ (left) in the ($m_{\textrm{Z}^{\prime}}$, $m_{\chi}$) plane and for 2HDM+a (right) as a function of model parameters: $\ma$, $\mA$, $\sintheta$, and $\tanbeta$ while fixing the values of other parameters, as indicated in the figures~\cite{pas-sus-23-012}. }
\label{fig:monoHiggs_tautau}
\end{figure}

The $M_{T}^{tot}$ distribution is used to extract the signal with a binned maximum likelihood fit. No significant excess of data above the SM background expectation is observed. Upper limits at the 95\% CL are set on model parameters for the baryonic-$\textrm{Z}^{\prime}$ model in a two-dimensional plane ($m_{\textrm{Z}^{\prime}}$, $m_{\chi}$) using a data set corresponding to 138 $\textrm{fb}^{-1}$ as shown in figure~\ref{fig:monoHiggs_tautau} (left). The interpretation for the 2HDM+a model is done for the first time in $\tau\tau$ channel with Run-2 data set corresponding to an integrated luminosity of 101 $\textrm{fb}^{-1}$. Figure~\ref{fig:monoHiggs_tautau} (right) shows exclusion limits at the 95\% CL as a function of the four model parameter - $\ma$, $\mA$, $\sintheta$, and $\tanbeta$.

\subsection{Search for mono dark Higgs ($b\bar{b}$)}
In the theoretical framework considered in this search~\cite{pas-sus-23-013}, the dark Higgs boson $\textrm{H}_{\textrm{D}}$ is the lightest state in the dark sector, and does not decay into DM particle $\chi$. Instead, it decays into SM particles by mixing with the SM Higgs boson $h$, with dominant branching fraction into a bottom quark-antiquark pair for $m_{\textrm{H}_{\textrm{D}}}$ below 135 GeV, and significant for masses up to 160 GeV. The mixing angle $\theta_{h}$ between the $\textrm{H}_{\textrm{D}}$ boson and the $h$ boson is set to 0.01, large enough to ensure prompt decay while small enough to have no observable effect on the couplings of the $h$ boson. This search targets events where DM particles are produced in association with a $\textrm{H}_{\textrm{D}}$ boson decaying into $b\bar{b}$ pair, using data set collected at $\sqrt{s}=13$ TeV during 2016--2018, corresponding to an integrated luminosity of 138 $\textrm{fb}^{-1}$. 

Events are required to have $\ptmiss>250$ GeV. The $b\bar{b}$ decay products are reconstructed into an anti-$k_{T}$ jet with distance parameter $R=1.5$, with $\pt>160$ GeV, and softdrop corrected mass $40<m_{\textrm{\textrm{SD}}}<300$ GeV. The $\textsc{DeepAK15}$ algorithm is used to identify large-area jets consistent with $b\bar{b}$ decays, and using a tagging score corresponding to a signal efficiency of 90--95\%. 

\begin{figure}[!h]
    \centering
    \includegraphics[width=0.32\linewidth]{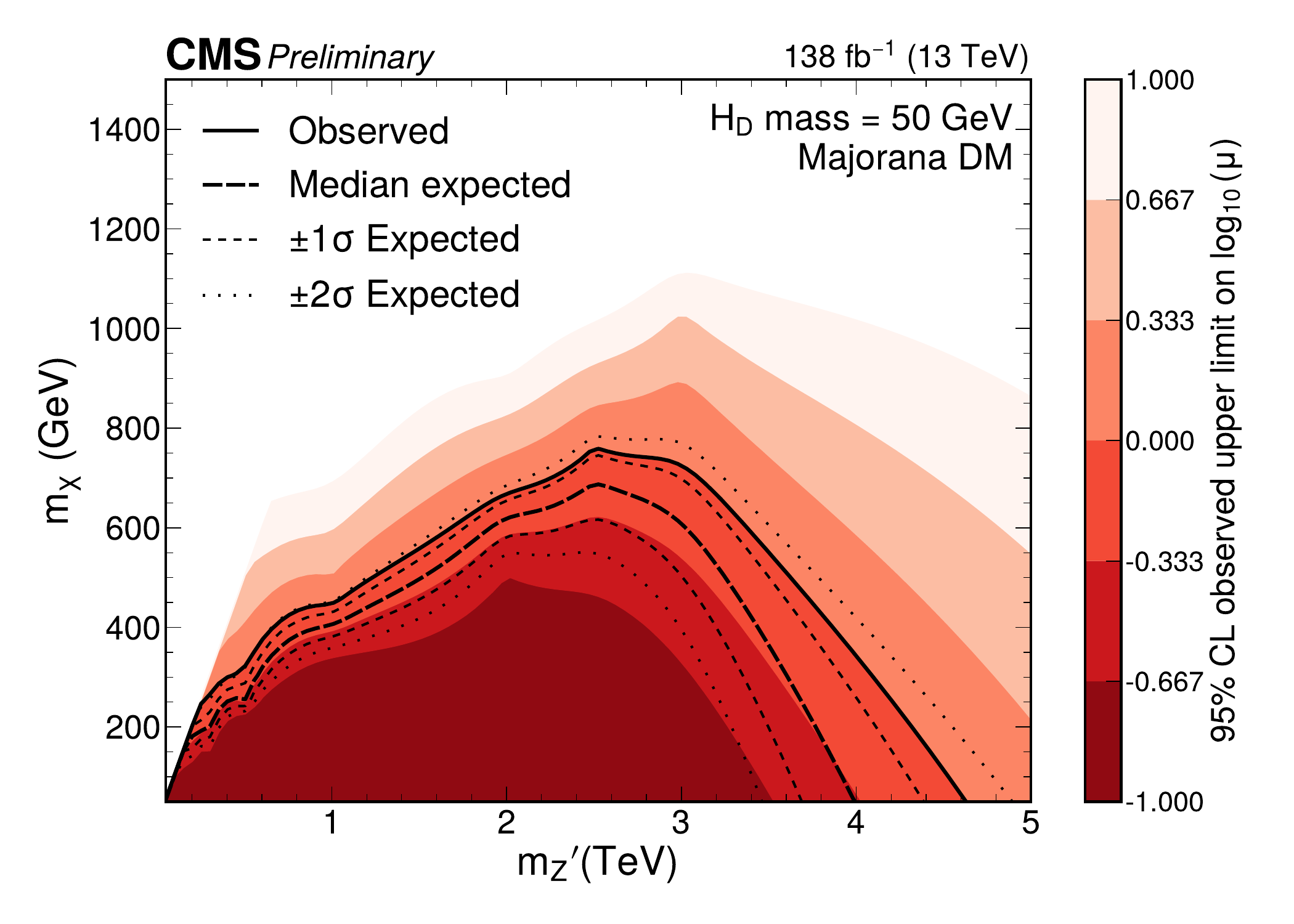}
    \includegraphics[width=0.32\linewidth]{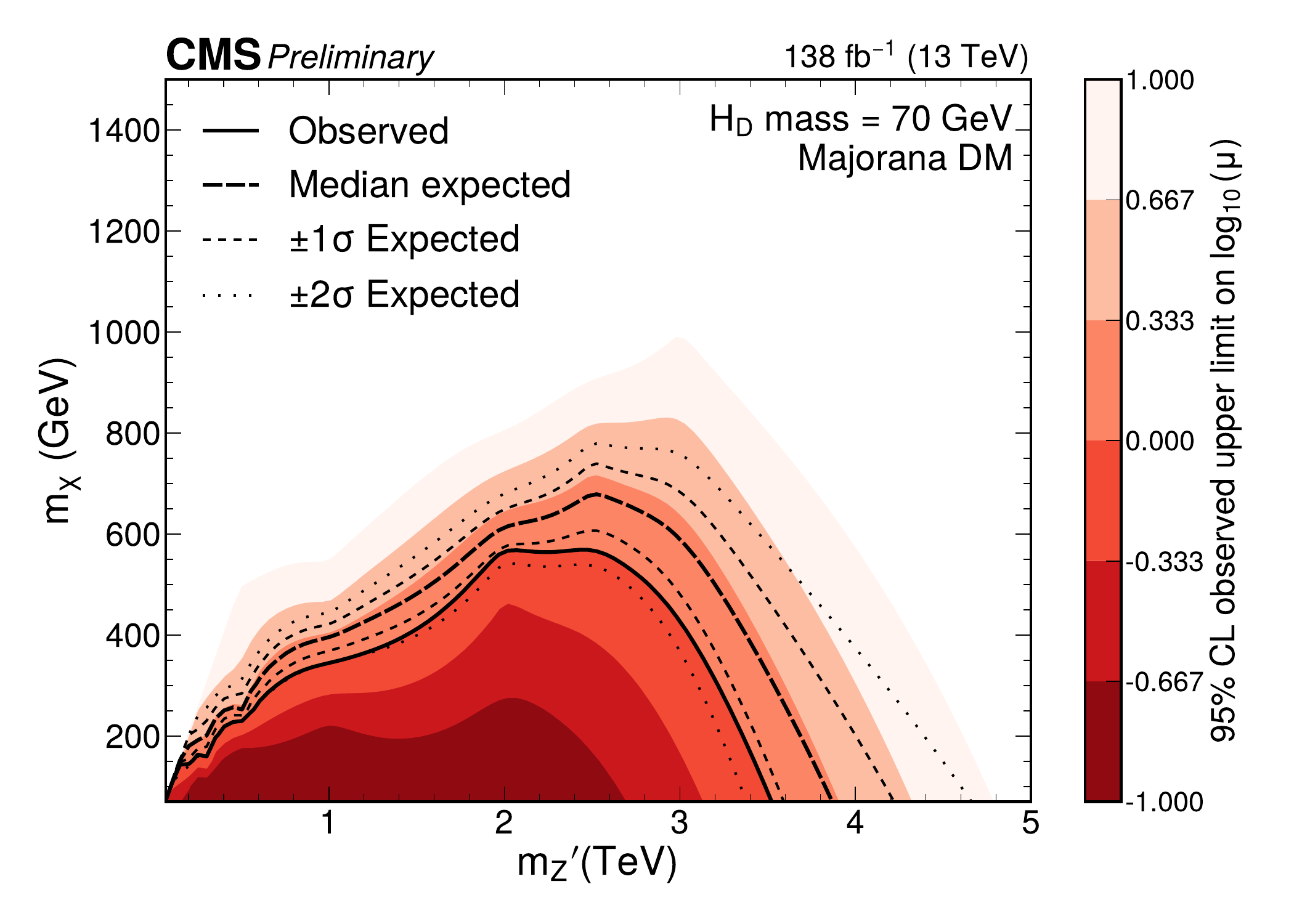}
    \includegraphics[width=0.32\linewidth]{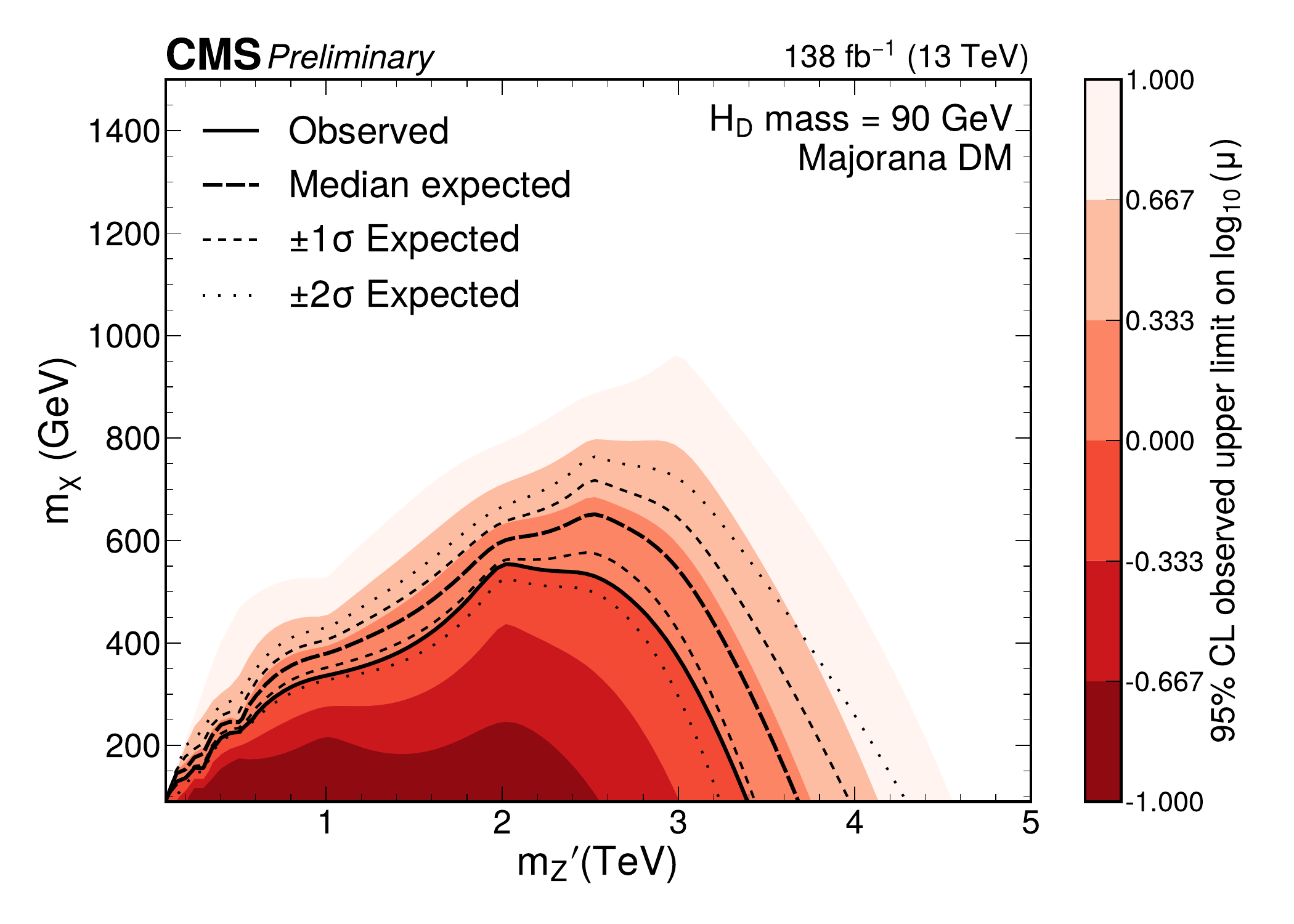}
    \includegraphics[width=0.32\linewidth]{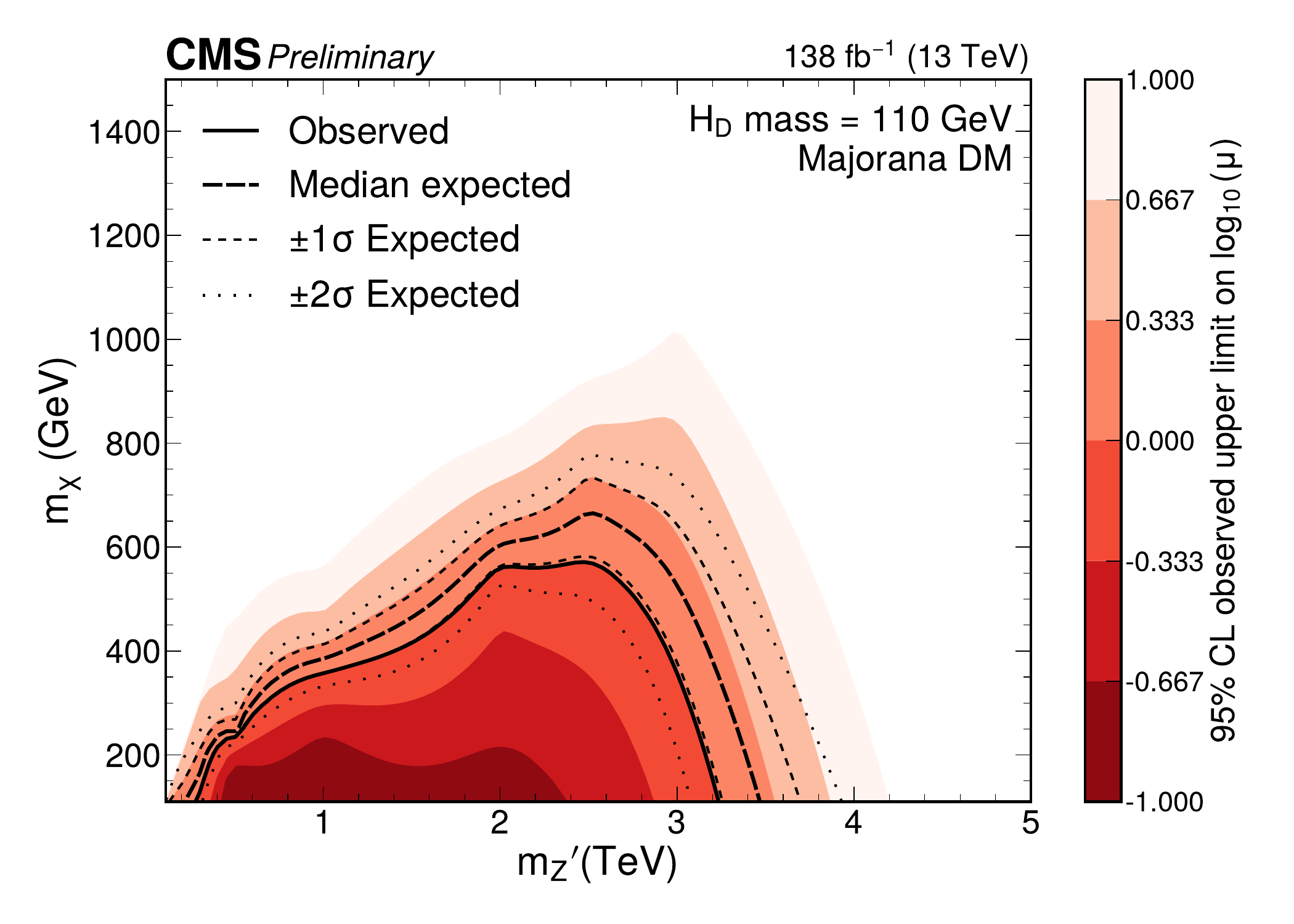}
    \includegraphics[width=0.32\linewidth]{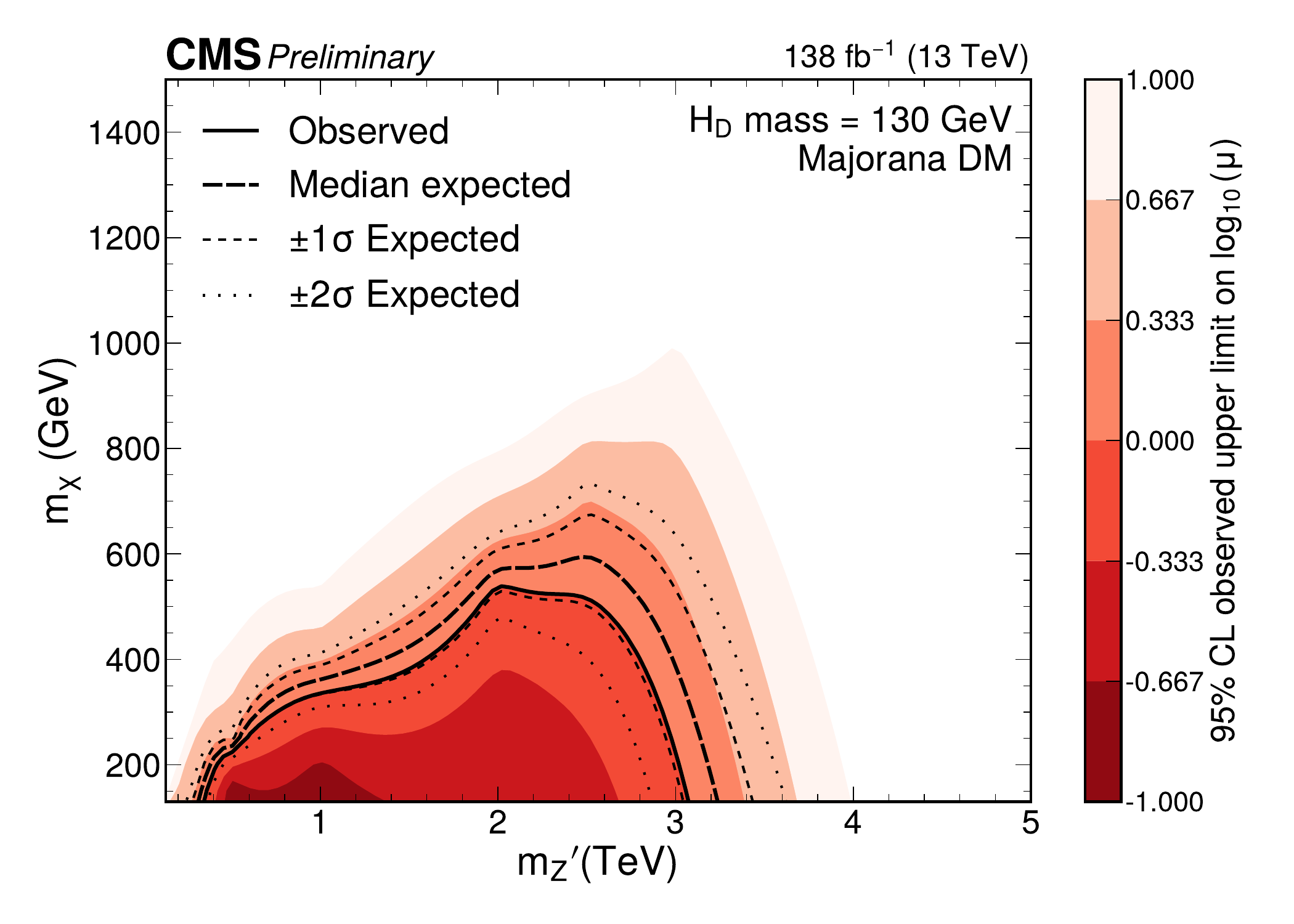}
    \includegraphics[width=0.32\linewidth]{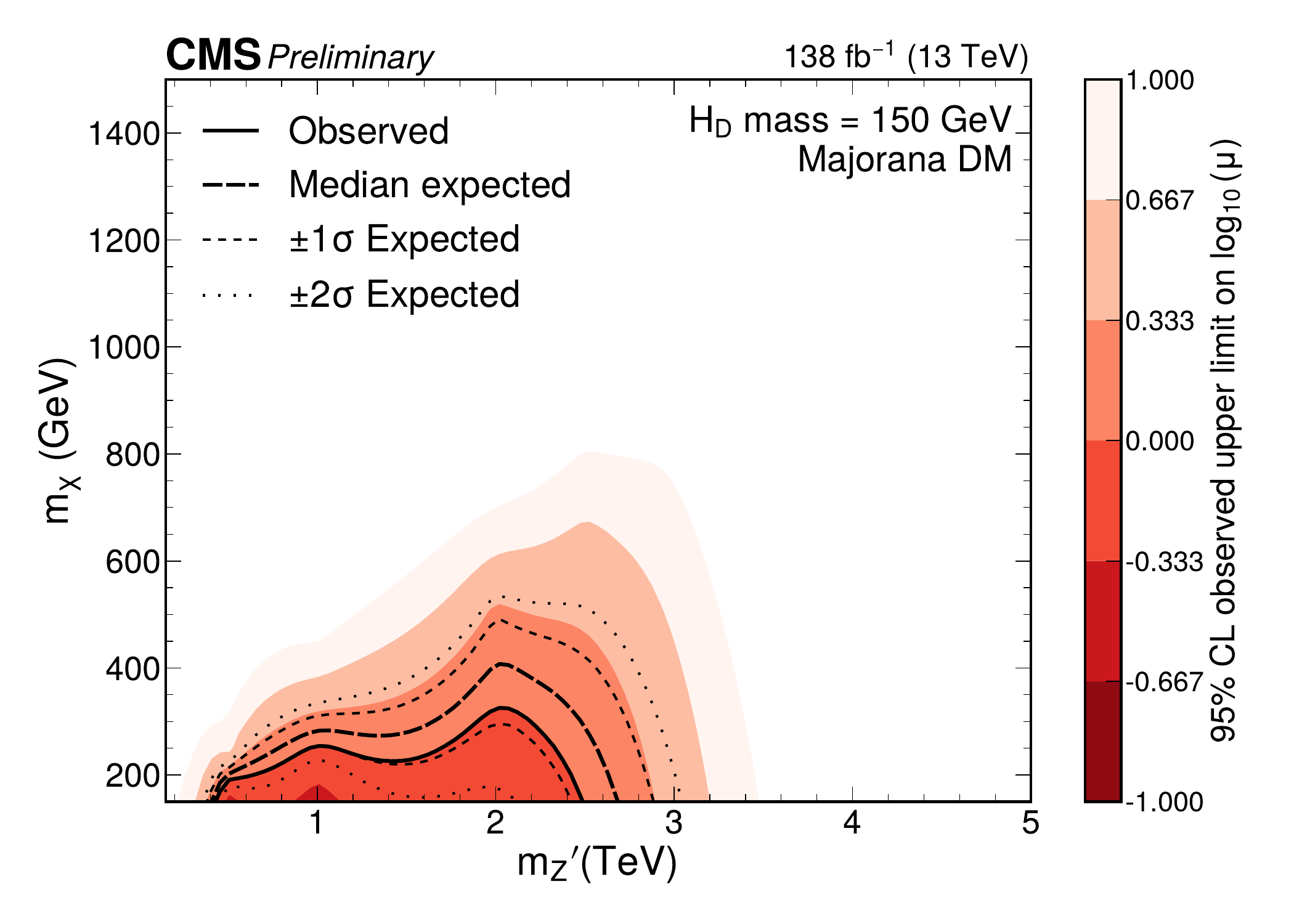}
    \caption{The expected and observed exclusion limits at the 95\% CL on the signal strength $\mu=\sigma/\sigma_{theory}$ as a function of $m_{\textrm{Z}^{\prime}}$ and $m_{\chi}$ for $m_{\textrm{H}_{\textrm{D}}}$ hypotheses of 50, 70, 90, 110, 130, and 150 GeV. Only scenarios with $m_{\chi} > m_{\textrm{H}_{\textrm{D}}}$ are considered. The black solid line indicates the observed exclusion boundary corresponding to $\mu=1$. The black dashed and dotted lines represent the expected exclusion and the 68 and 95\% CL intervals around the expected boundary, respectively. Parameter combinations corresponding to larger values of $\mu$ are excluded~\cite{pas-sus-23-013}.}
    \label{fig:monodarkHiggs_bb}
\end{figure}

No significant excess above the SM background predictions is observed in data, and exclusion limits on signal strength $\mu$ are placed for different signal hypotheses. The coupling strength of the mediator with SM quarks and with DM particles is set to 0.25 and 1, respectively. Exclusion limits are calculated in two-dimensional parameter space of the DM and mediator masses, considering scenarios where $m_{\textrm{DM}} > m_{\textrm{H}_{\textrm{D}}}$. Figure~\ref{fig:monodarkHiggs_bb} shows the exclusion limits at 95\% CL for different $m_{\textrm{H}_{\textrm{D}}}$ hypotheses ranging from 50 GeV to 150 GeV. The darker shade corresponds to a smaller upper limit, i.e., more stringent constraints. Values of the $m_{\textrm{Z}^{\prime}}$ up to 2.5--4.5 TeV are excluded depending on $m_{\textrm{H}_{\textrm{D}}}$. Limits for $m_{\textrm{H}_{\textrm{D}}}<160$ GeV hypotheses are set for the first time with CMS data.

\section{Summary and Outlook}
CMS has a rich and broad search program for dark matter. The recent results showcased in this report probe the role of the Higgs boson as a potential portal between the SM and dark matter with Run-2 data. So far, the observed data agree with the SM predictions with no hint of new physics. With the enormous dataset in Run-3 and the upcoming HL-LHC, improved analysis techniques along with upgraded detectors will enhance the physics reach.


\begin{thebibliography}{99}
\bibitem{cmspaper}{CMS Collaboration, JINST 3, S08004 (2008), \href{https://iopscience.iop.org/article/10.1088/1748-0221/3/08/S08004}{10.1088/1748-0221/3/08/S08004}.}

\bibitem{pas-sus-24-007}{CMS Collaboration, CMS-PAS-SUS-24-007, 2025, \href{https://cds.cern.ch/record/2931051}{cds.cern.ch/record/2931051}}

\bibitem{pas-sus-23-012}{CMS Collaboration, \emph{Search for dark matter produced in association with a Higgs boson decaying to a $\tau$ lepton pair in proton-proton collisions at $\sqrt{s}=13$ TeV} (2025), \href{https://doi.org/10.1007/JHEP10(2025)170}{10.1007/JHEP10(2025)170}}

\bibitem{pas-sus-23-013}{CMS Collaboration, CMS-PAS-SUS-23-013, 2025, \href{https://cds.cern.ch/record/2940270}{cds.cern.ch/record/2940270}}
\end{thebibliography}
\end{document}